\documentclass[twocolumn,amsmath,amssymb]{revtex4}

\def\gsim{ \lower .75ex \hbox{$\sim$} \llap{\raise .27ex
\hbox{$>$}} }
\def\lsim{ \lower .75ex \hbox{$\sim$} \llap{\raise .27ex
\hbox{$<$}} }

\usepackage{graphicx}
\usepackage{dcolumn}
\usepackage{bm}
\usepackage{graphicx}

\usepackage{amssymb}
\usepackage{bbm}
\def\ba{\begin{eqnarray}}
\def\ea{\end{eqnarray}}
\def\be{\begin{equation}}
\def\ee{\end{equation}}
\def\ben{\begin{equation} \nonumber}
\def\een{\end{equation}}
\def\baray{\begin{eqnarray*}}
\def\earay{\end{eqnarray*}}

\def\SE{{S_{\mbox{\tiny E}}}}
\def\SB{{S_{\mbox{\tiny B}}}}
\def\SBB{{S_{\mbox{\scriptsize B}}}}

\begin{document}

\title{Stability in and of de Sitter space}

\author{Benjamin Shlaer}

\affiliation{Institute of Cosmology, Department of Physics and Astronomy, \\
Tufts University, Medford, MA  02155, USA }

\begin{abstract}
\noindent We demonstrate that possession of a single negative mode is not a sufficient criterion for an instanton to mediate exponential decay.  For example, de Sitter space is generically stable against decay via the Coleman-De Luccia instanton.  This is due to the fact that the de Sitter Euclidean action is bounded below, allowing
for an approximately de Sitter invariant false vacuum to be constructed. 
\end{abstract}

\maketitle
\noindent
{\bf A warm-up with quantum mechanics.}
It is well known that certain quantum mechanical states are meta-stable, an example of which is illustrated  in Fig.~\ref{potential1} below.  (We assume the reader is familiar with the seminal review by S. Coleman \cite{Coleman:1978ae}.)
 \begin{figure}[h] 
   \centering
   \includegraphics[width=3.0in]{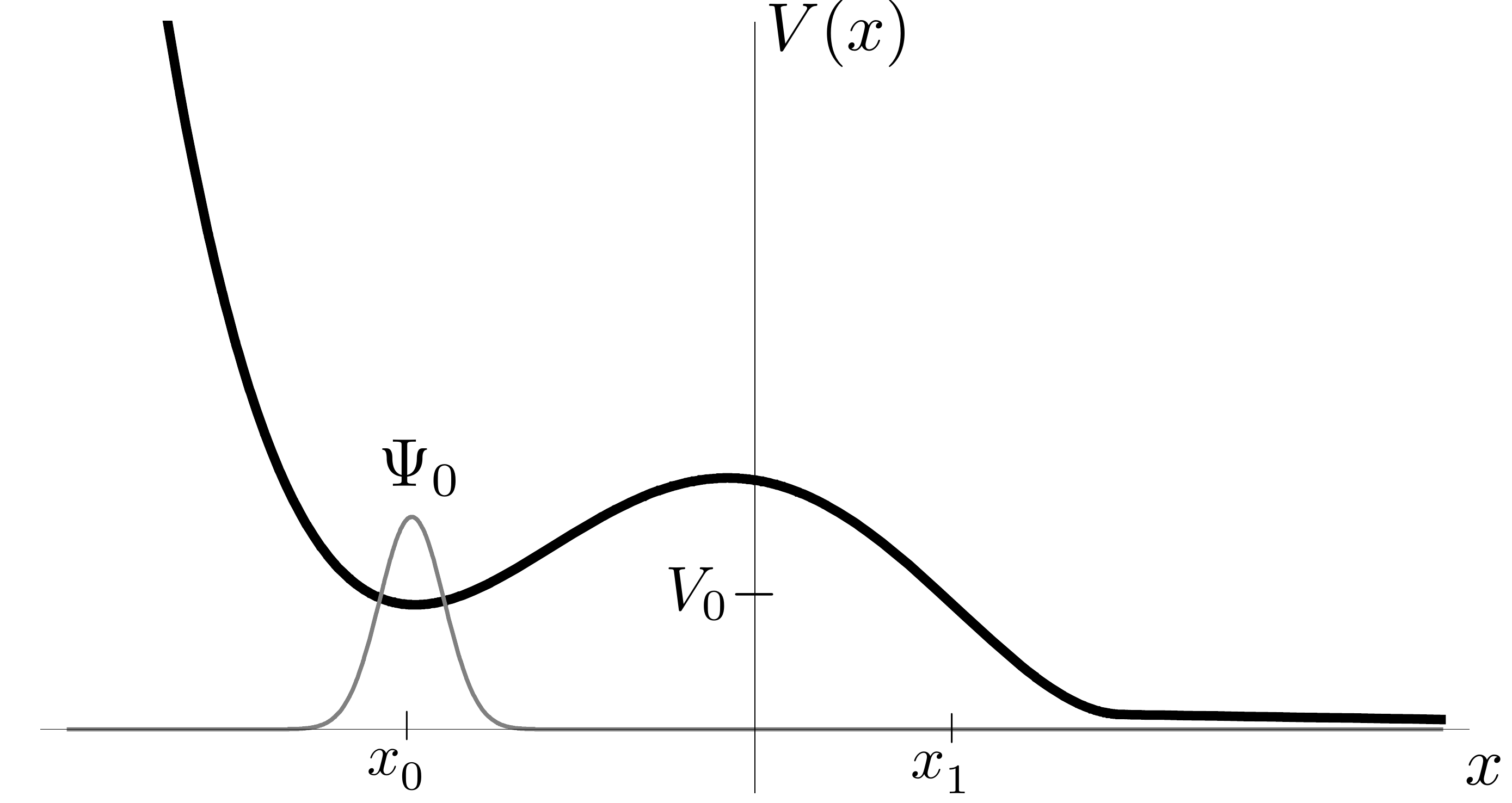} 
   \caption{A quantum mechanical potential exhibiting exponential decay: $1-P_{\rm decay} = \left|\left<\Psi_0|\exp(-iHt)|\Psi_0\right>\right|^2 \approx \exp(-\Gamma|t|).$} 
   \label{potential1}
\end{figure}
We can calculate the decay rate using the Euclidean path integral formalism:
\ba\label{eqn:pathintegral}
\left<x_0|e^{- H T}|x_0\right>&=& \int D[x]e^{-\SE[x]}.
\ea
Evaluating the Euclidean action for the simplest non-trivial instanton $x_{\rm B}$ is equivalent to integrating $\int p\,dx$ across the barrier and back.  Thus `the bounce' action is
\ba
\SBB &=& 2\int_{x_0}^{x_1} dx\sqrt{2m(V(x)-V_0)},
\ea
and the decay rate is twice the dilute instanton gas correction to the perturbative ground state energy,
\be
E_{\rm fv}^{\mbox{\scriptsize nonpert.}} \sim -\frac{i}{4}\nu_{\rm fv}^{-1}\exp(-\SBB),
\ee
where the perturbative density of states $\nu_{\rm fv} = (\Delta E_{\rm fv})^{-1}$.
The decay rate is thus
\be\label{eqn:qmdecay}
\Gamma \,\,\sim \,\,\tfrac{1}{2}\nu_{\rm fv}^{-1}e^{-\SB}.
\ee

\begin{figure}[h] 
   \centering
   \includegraphics[width=3.0in]{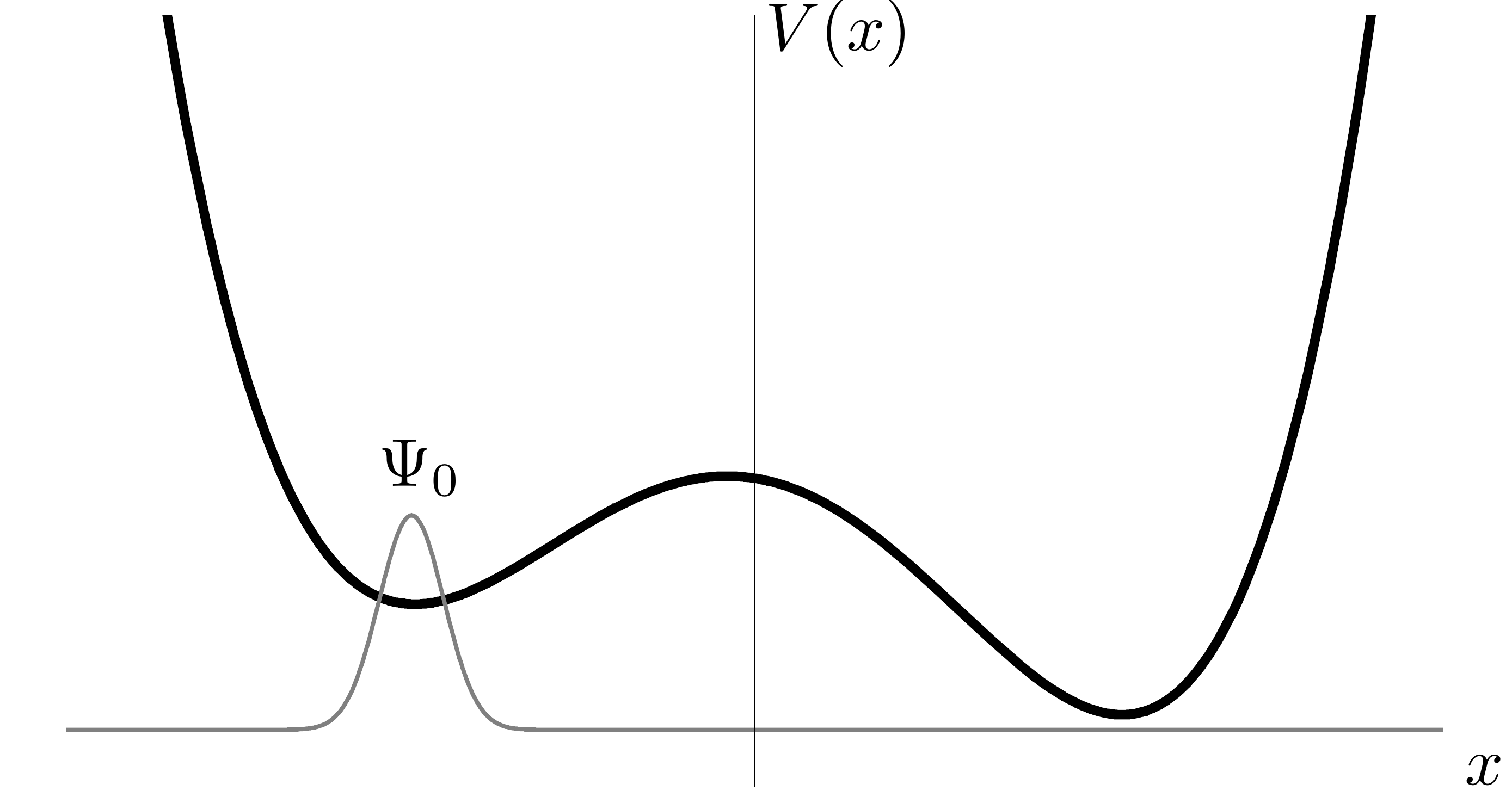} 
   \caption{A quantum mechanical potential with a stable false vacuum $\left|\Psi_0\right>$.  We assume $e^{-\SB} \ll 1$.} 
   \label{potential2}
\end{figure}

We have preserved a didactic factor of $1/2$ in the decay rate for illustrative purposes later on.  The semi-classical approximation allows the path integral to be evaluated in the Gaussian approximation, i.e., the decay rate is proportional to the square root of a functional determinant, except care must be taken with the negative and zero eigenvalues. 
It is the famous single negative mode which is responsible for the factor of $i/2$, and thus exponential decay.
Heuristically speaking, the energy of the false vacuum $\left|\Psi_0\right>$ has nonzero imaginary part.  
More precisely, if the
false vacuum is normalizable, and the energy eigenstates are not, the false vacuum cannot be thought of as an approximate energy eigenstate.  
Such an initial state will certainly decay.
However, if the false vacuum $\left.|\Psi_0\right>$ is spectrally composed of the {\em discrete} portion of the Hamiltonian $H$, exponential decay cannot occur.  
Absent a symmetry, the generic probability for the false vacuum to decay at all is proportional to the ratio of the (accessible) densities of state:
\be
\limsup_{t\to\infty} \frac{P_{\rm decay}}{1-P_{\rm decay}} \sim \frac{\nu_{\rm tv}(E_{\rm fv})}{\nu_{\rm fv}(E_{\rm fv})}{\rm e}^{-S_{\rm B}}.
\ee
A false vacuum which is, intuition notwithstanding, stable \cite{Nieto:1985ws} is shown in
the potential of Fig.~\ref{potential2}.

By changing a portion of the potential far away from `the bounce' $x_{\rm B}$, we have stabilized $|\Psi_0\rangle$.  This false vacuum can now be thought of as an approximate energy eigenstate.  In the Euclidean formalism, this is because despite having a negative mode locally, the path integral is convergent, and so only real contours are integrated over.  

 \begin{figure}[h] 
   \centering
   \includegraphics[width=3.2in]{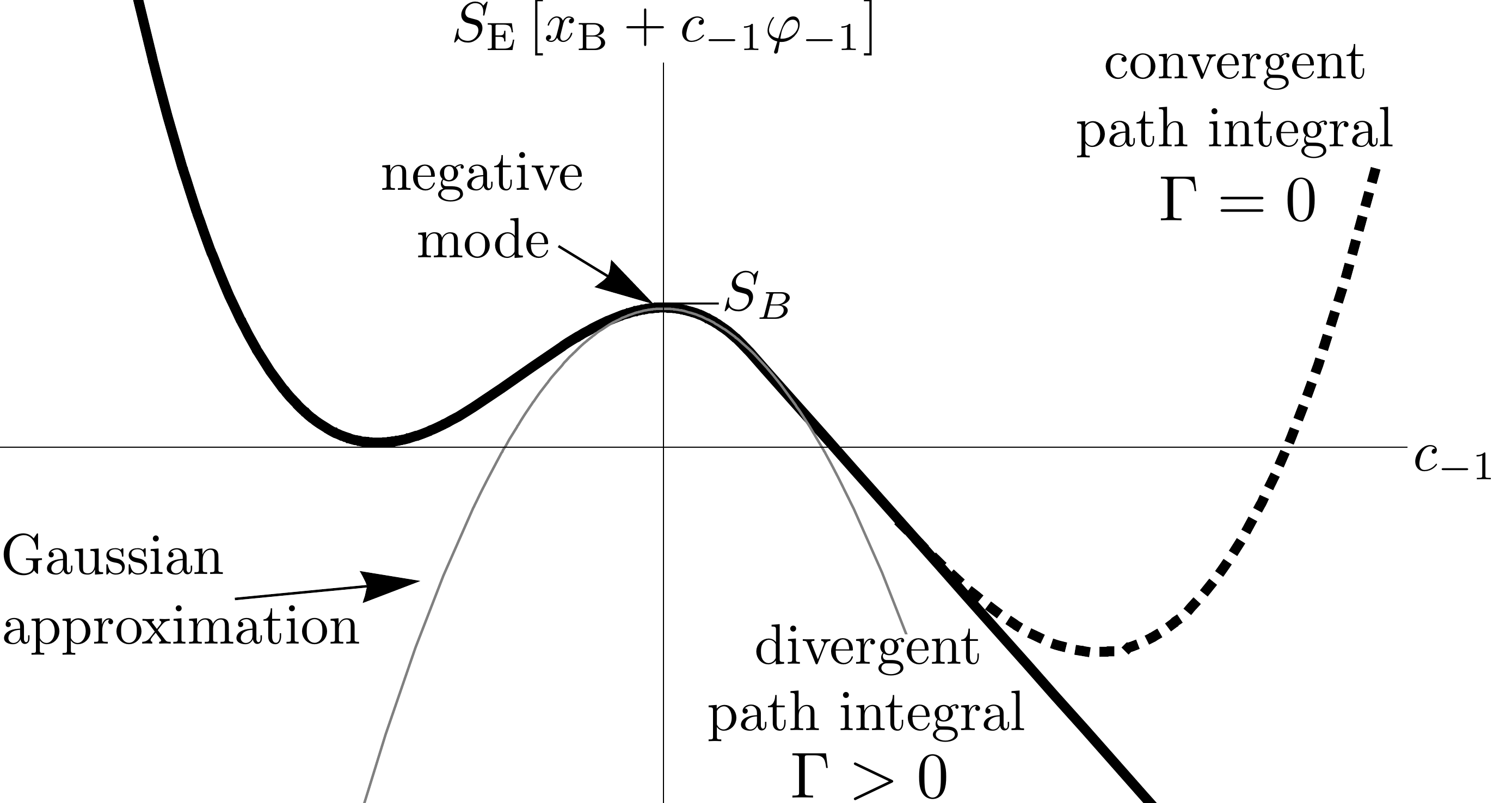} 
   \caption{The Euclidean action possesses a single negative mode.  This is meaningful if it signifies a divergence in Eq.~(\ref{eqn:pathintegral}).  Sagredo's Gaussian approximation for $\Gamma$ misses a factor of $\tfrac{1}{2}$ (solid curve) and $\tfrac{0}{2}$ (dashed curve) for the potentials in Figs.~\ref{potential1} and \ref{potential2}, respectively.} 
   \label{action}
\end{figure}

A convenient basis for the space of Euclidean paths is given by $\varphi_\lambda$, the eigensolutions to the second functional variation of the action about the one-instanton solution $x_{\mbox{\scriptsize B}}$.  It is here where the single negative mode is manifest, and we plot the Euclidean action for the above two potentials in Fig.~\ref{action} below.  The same instanton has vastly different interpretations for the two potentials.  One is responsible for the instability of exponential decay, and the other is a (typically) inconsequential contribution to the energy splitting amongst the various bound states.

This is precisely the reason why Sagredo \cite{Coleman:1978ae} misses the factor of one-half in the calculation of Eq.(\ref{eqn:qmdecay}).  His na\"ive analytic continuation assumes the path integral is divergent due to the integration over the single negative mode along both
the negative and positive real axes, when in fact it is only the positive real axis contour which must be moved;  As the bounce amplitude is increased to the right in Fig.~\ref{potential1} past the classical turning point $x_1$, the particle spends more Euclidean time at increasingly negative potential energy, manifesting the divergence we associate with the negative mode.  But no divergence occurs for the integral over the negative real axis, hence the unexpected factor of $\tfrac{1}{2}$ in the decay rate.  In the case shown in Fig.~\ref{potential2}, both directions are convergent.  The contour is kept on the real axis, giving no imaginary contribution to the energy and hence no exponential decay.  This mechanism has been experimentally observed for excited atoms in reflecting cavities \cite{Martini}.

Although not always manifest within the bounce formalism, it is possible to fine-tune the potential in Fig.~\ref{potential2} to destabilize $\left|\Psi_0\right>$ via resonance.  This occurs when the energy splitting between the two perturbative (left and right) states becomes of order the tunnel coupling.
It should be emphasized that without a symmetry to protect the degenerate energies of the two perturbative states, this can only be achieved with a fine tuning of order $e^{\SB/2}$.  One example of such a symmetry is the ${\mathbbm Z}_2$ parity symmetry $V(x) = V(-x)$.  A fine-tuned example of resonance is the Salpeter-Hoyle triple-alpha process of carbon production in stars.

The transition between stable and meta-stable occurs with the appearance of an exponential hierarchy between the true- and false- vacuum density of accessible states, i.e. states not forbidden by conservation laws.

The appearance of a saddle point of the Euclidean action is a sign that the path integral is either divergent (in which case the contour is moved, effectively creating a local minimum in the pushed coordinates) or yielding a subdominant contribution.  The path integral is {\em correctly} defined via analytic continuation if and only if it is necessary to do so, i.e., if the contour on the real-axis gives a divergent answer.  
The divergence from the negative mode implies the spectrum of the unstable state in question lives in the continuous portion of the Hamiltonian, a necessary but not sufficient condition for exponential decay.

\begin{figure}[h] 
   \centering
   \includegraphics[width=2.6in]{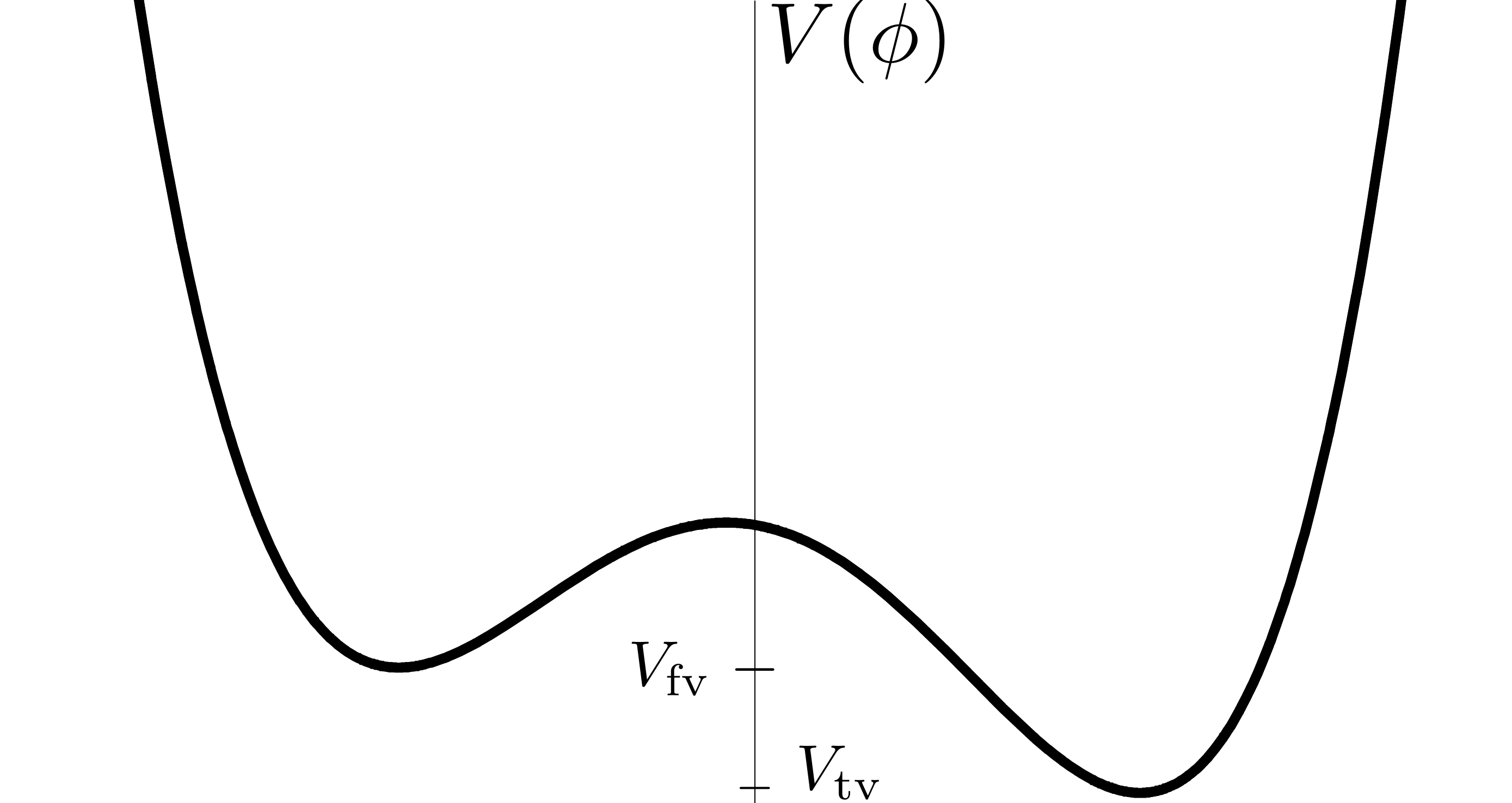} 
   \caption{A potential for the field $\phi$ which exhibits false vacuum decay in flat  (and perhaps AdS) space.} 
   \label{potential3}
\end{figure}

{\bf  Field theory.} 
If we move on to field theory in flat space, we can now find exponential decay \cite{coleman} from the potential in Fig.~\ref{potential3}.  Spherical solitonic domain walls interpolate between the two vacua, each bubble being in possession of a single negative mode.
The essential point here is that the negative mode actually leads to a divergence in the path integral, and the
false vacuum energy density acquires an imaginary contribution from instantons.  As before, there is a subtle factor of $1/2$ in the decay rate per unit volume, since making the Euclidean $S^3$ bubble radius arbitrarily smaller than the critical value does {\em not} result in a divergence.  The only divergence occurs when the $S^3$ radius is made arbitrarily large.  
A similar argument applies at finite temperature \cite{Linde:1981zj}; Decay is guaranteed.

If we compactify all three spatial dimensions, the infrared divergence of the Euclidean path integral is cured, along with the associated instability.  
While resonances \cite{Tye:2009rb} are possible, finite volumes lack the continuum modes necessary for exponential decay.  In practice, a density of states exponentially dependent on volume is coupled to the false vacuum by bubble collisions, yielding a sharp transition between stable and meta-stable false vacuum as a function of spatial volume --- decay remains generic.

{\bf The Coleman-De Luccia instanton.} 
The QFT Euclidean path integral in de Sitter space resembles that of compactified flat space in that
there are no infrared divergences.  The thin-walled Coleman-De Luccia \cite{cdl}  instanton in de Sitter space describes
an $S^3$ domain wall dividing a four-sphere into two distinct constant-curvature regions.  The interior region is approximately a homogenous configuration of $\phi$ in the right-hand well of Fig.~\ref{potential3}, and the exterior (which is also topologically a three-ball) has $\phi$ localized near the bottom of the left-hand well.  While this has the required single negative mode corresponding to dilatation of the bubble, it is clear from the compact topology of the background that the path integral is infrared convergent.  Hence the negative mode is just a sign that the global minimum has not been expanded about.  In de Sitter space, Sagredo would have missed a salient factor of zero-halves in the decay rate.  

None of our conclusions change when we consider the thick-walled CDL instanton, including the extremal case known as the Hawking-Moss instanton \cite{Hawking:1981fz}.  This is a homogenous field configuration with $\phi$ at the local maximum of $V(\phi)$ shown in Fig.~\ref{potential3}.  It is known that the HM instanton possesses a single negative mode precisely when the CDL instanton ceases to exist \cite{Gratton:2000fj}.  From the figure, it is clear that this negative mode cannot lead to a divergence as long as $V(\phi)$ is bounded below.

{\bf Density of states.}  
The infrared-finite property of quantum corrections to classical vacua still allows for very different behaviors at finite volume in quantum mechanics compared to field theory.  
While true exponential decay is excluded in both systems, the typical field theory easily loses its stable false vacua because of the exponentially large ratio of densities of state.

In 4d de Sitter space, the false vacuum has $O(4)$ symmetry, and in particular, $O(3)$ symmetry about any observer.  The accessible states are those which respect this large symmetry.  In other words, when a bubble nucleates, it grows and exits
the static patch without exciting the many modes of the QFT true vacuum.  For this reason, the ratio of the accessible densities of state resembles
that of one-dimensional quantum mechanics;  Decay is a one-degree-of-freedom process.  We can calculate $P_{\rm decay}$ of $4$-dimensional de Sitter space
by reducing it to $1+1$-dimensions, where there is no difference between the density of states and the {\em accessible} density of states.

In the static patch, the ratio of accessible densities of state comes from the minisuperspace entropy difference, a Gibbons-Hawking-like $\Delta S_{\mbox{\scriptsize `GH'}} \sim \log(\ell^4 \Delta V)$, where $\Delta V = V_{\rm fv} - V_{\rm tv}$ and $\ell$ is the de Sitter radius.
Hence, the ratio of the accessible densities of state is only polynomial in $\ell$, and indeed $P_{\rm decay} \sim \ell^4\Gamma_{\rm CDL}\ll 1$.  This is our main result.


{\bf  Canonical quantization methods.}
We can investigate vacuum decay from a canonical point of view by considering the massive
bosonic Schwinger model, consisting of a charged scalar field in $1+1$ dimensions \cite{Garriga:1994bm}.  Here, the fundamental particles replace the solitonic domain walls, and a Poincar\'e (or  de Sitter) invariant electric field plays the role of vacuum energy.  Gravity is trivial in two dimensions, so we choose the geometry by hand.  In flat space, a false vacuum can be constructed with a  homogenous electric field $\vec{E}_0 =  g$, where $g$ is the quark charge.  Schwinger pair production is the mechanism for false vacuum decay, through which the electric field decreases in units of  $g$ via $\langle \vec{E}(t)\rangle = \vec{E}_0\exp(-\Gamma t^2/2).$

An interesting thing happens if we compactify the Schwinger model with $\vec{E}_0 \approx g$ on a circle.  In finite volume, the Hamiltonian becomes bounded below.  When a pair is produced to neutralize the interceding field strength, it accelerates apart, but then crosses on the opposite side of the circle.
This crossing further reduces the field strength to $\approx -g$, so the pair now decelerates, turns around, and returns to the starting configuration; The decay products of the false vacuum are excited mesons.  Because the spectrum of these bound states is discrete, they are only produced by resonances with the false vacuum.  Hence without fine-tuning, the false vacuum is stable for the compact Schwinger model.  This stability breaks down when the spatial volume becomes exponentially larger than the size of an instanton.  This is because the one-meson density of states becomes large enough to guarantee resonance.  We can ignore multiple-meson states, which carry additional suppression factors of $\exp(-\SBB)$.
The generalization of this to four dimensions is to put the Brown-Teitelboim \cite{Brown:1988kg} model on a spatial three-sphere, a system that has stable false vacua for sufficiently small volume.
{\em Meaningful results can only be obtained by including the back-reaction of
the domain walls on the field strength.}  In the absence of the back-reaction, the Hamiltonian is unbounded below.  The differences between the free and back-reacting theories is particularly stark at finite volume.  This can be demonstrated in the path-integral formalism as well.  Only when the back-reaction is taken into account is the Euclidean path integral convergent on the cylinder, as well as on compact Euclidean geometries.  The probe approximation fails to capture important physics.  

\begin{figure}[h] 
   \centering
   \includegraphics[width=3.4in]{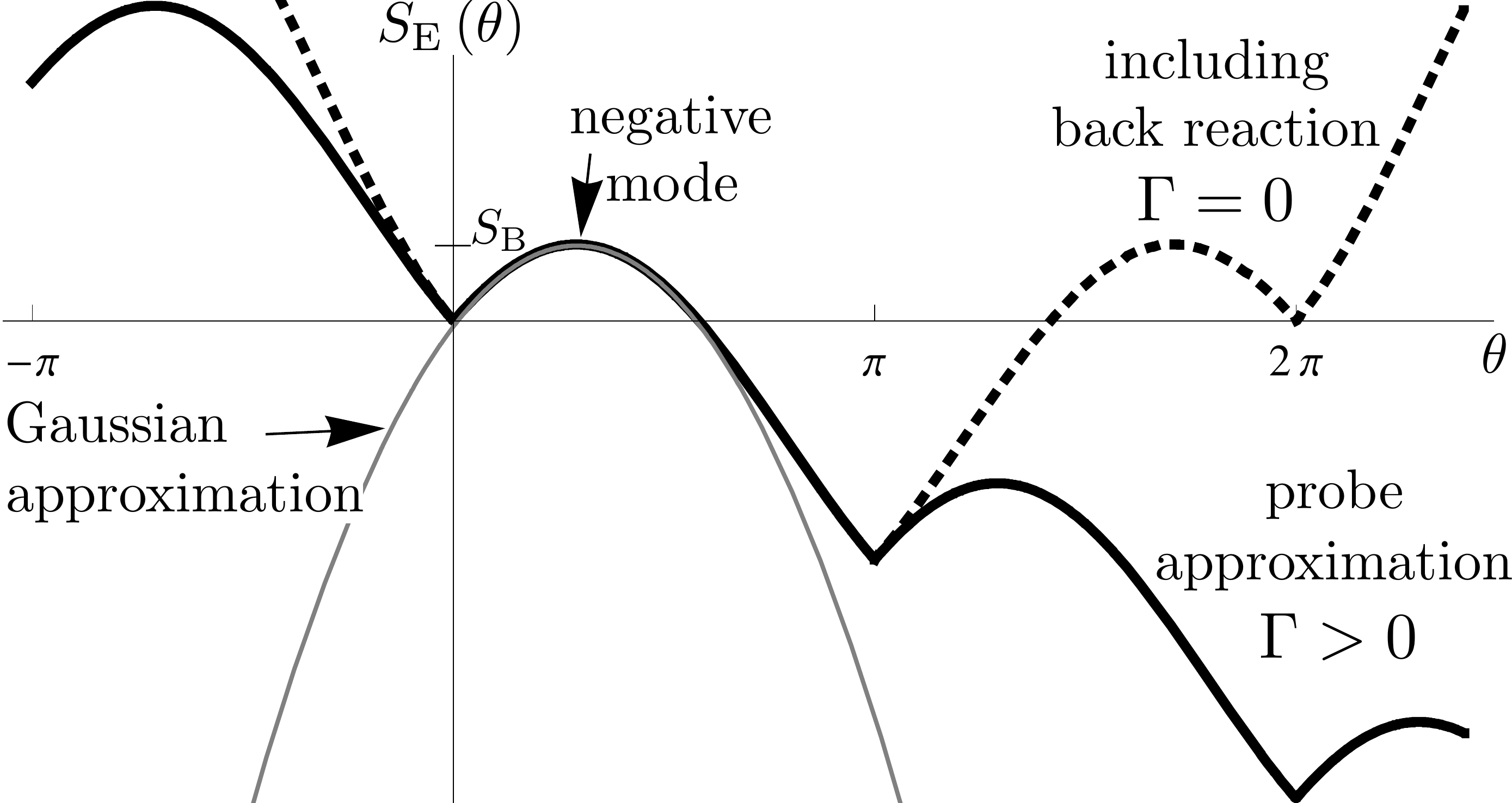} 
   \caption{The Euclidean action is plotted as a function of $S^1$ bubble wall radius.  The coordinate $\theta$ is the polar angle on $S^2$ whose fundamental domain is $\left[0,\pi\right]$.  This is extended due to the multivaluedness of the electric field as a function of bubble configuration.  The probe approximation (solid curve) gives qualitatively different results, and is thus invalid.} 
   \label{cdlaction}
\end{figure}

The probe approximation has previously been used \cite{Garriga:1994bm} to argue in favor of pair production in the two-dimensional de Sitter-Schwinger model.  In these analyses the theories are free, and the Hamiltonians are unbounded below.  Although this is amenable to the use of Bogolyubov coefficients, which predict vacuum decay in the sense of a small, constant physical number density of nucleated charged particles, the previously mentioned faults of the probe approximation render such results suspect.  If one neglects the back-reaction, the action can be arbitrarily lowered by growing a Euclidean bubble and contracting it about the antipodal point, re-inflating it with opposite orientation, contracting it about the original point, etc.  This is illustrated in Fig.~\ref{cdlaction}.

{\bf Discussion}
The preferred de Sitter-invariant quantum state is the Hartle-Hawking state.  On one static patch this is described by a thermal density matrix $\rho \propto \exp( - 2\pi \ell H)$.  For the false vacuum, we construct this using the perturbative action.  We argue that the non-perturbative corrections to this state are negligible, and so de Sitter invariance, including time-translation invariance, is preserved.  There are stable {\em perturbative} Hartle-Hawking states for both minima of the potential, and they are well approximated by deleting certain entries of the {\em exact} Hartle-Hawking density matrix, a procedure which preserves time-independence. Because each perturbative vacuum has an infinite density of states, this method of constructing false vacua
is compatible with cluster decomposition; Such states closely resemble their perturbative counterparts.

Although this is a thermal system, it cannot be said that nucleation of a real bubble will eventually occur.  To quantify this, we state our result as follows:  The expectation value of the vacuum energy {\em as measured by an inertial observer in the de Sitter false vacuum} is time-independent, and equal to
\be
\left<\,\rho_{\rm vac}\,\right> = \Delta V e^{-\Gamma_{\rm CDL} \ell^4} + V_{\rm tv} \approx V_{\rm fv},
\ee
where $\Gamma_{\rm CDL}$ is the na\"ive decay rate per unit volume. 
This is in stark contrast to the result in flat or AdS space, where an ensemble of observers in the false vacuum finds the usual time-dependent result,
\be
\left<\,\rho_{\rm vac}\,\right> = \Delta Ve^{-\Gamma_{\rm CDL} V_4(t)} + V_{\rm tv},
\ee
where $V_4(t)$ is the 4-volume of the past light cone of the observer since initial conditions.
A false vacuum in de Sitter space is a stable one in the same way as the
false quantum mechanical vacuum of Fig.~\ref{potential2}.  

We have provided evidence that de Sitter false vacua
are stable when $\Gamma_{\rm CDL} \lesssim \ell^{-4}$, which is generically the case.  If one takes the Gibbons-Hawking entropy seriously, then decay of de Sitter space will involve a dynamical enlargement of the Hilbert space \cite{banks}, a task we are happy to avoid.
When the lifetime of the false vacuum is less than a Hubble time, the Coleman-De Luccia instanton does indeed mediate exponential decay.  Since generically $P_{\rm decay} \ll 1$, the dynamics of the landscape \cite{landscape} requires re-evaluation.

{\bf Acknowledgments.}
We thank Tom Banks, Jose Blanco-Pillado, Adam Brown, Larry Ford, Jaume Garriga, Matthew Johnson, Ferdinand Kuemmeth, Alberto Nicolis, Ken Olum, Henry Tye, Alex Vilenkin, and Erick Weinberg for helpful conversations.  Support was received through Foundational Questions Institute grant RFP2-08-26A.

\end{document}